\documentclass[pra,showpacs,twocolumn,groupedaddress]{revtex4}

\usepackage{amsfonts}
\usepackage{psfrag}
\usepackage{graphicx}

\newtheorem{theorem}{Theorem}[section]

\newtheorem{proposition}[theorem]{Proposition}
\newtheorem{corollary}[theorem]{Corollary}
\newtheorem{definition}[theorem]{Definition}

\newcommand{\qed}{\nobreak \ifvmode \relax \else
      \ifdim\lastskip<1.5em \hskip-\lastskip
      \hskip1.5em plus0em minus0.5em \fi \nobreak
      \vrule height0.75em width0.5em depth0.25em\fi}

\begin{document}

\title{Adiabatic approximation in open quantum systems}
\author{M.S. Sarandy}
\email{msarandy@chem.utoronto.ca}
\author{D.A. Lidar}
\email{dlidar@chem.utoronto.ca}
\affiliation{Chemical Physics Theory Group, Department of Chemistry, and
Center for Quantum Information and Quantum Control, University of
Toronto, 80 St. George St., Toronto, Ontario, M5S 3H6, Canada} 

\begin{abstract}
We generalize the standard quantum adiabatic approximation to the case of
open quantum systems. We define the adiabatic limit of an open quantum
system as the regime in which its dynamical superoperator can be decomposed
in terms of independently evolving Jordan blocks. We then establish validity and
invalidity conditions for this approximation and discuss their applicability
to superoperators changing slowly in time. As an example, the adiabatic
evolution of a two-level open system is analyzed.
\end{abstract}

\maketitle

\section{Introduction}

The adiabatic theorem~\cite{Born:28,Kato:50,Messiah:book} is one of the oldest and most
useful general tools in quantum mechanics. The theorem posits, roughly, that
if a state is an instantaneous eigenstate of a sufficiently slowly
varying
Hamiltonian
$H$ at one time, then it will remain an eigenstate at later times, while its
eigenenergy evolves continuously. Its role in the study of slowly varying
quantum mechanical systems spans a vast array of fields and applications,
such as energy-level crossings in molecules \cite{Landau:32,Zener:32},
quantum field theory \cite{Gellmann:51}, and geometric phases \cite%
{Berry:84,Wilczek:84}. In recent years, geometric phases have been proposed
to perform quantum information processing~\cite{ZanardiRasseti:99,ZanardiRasseti:2000,Ekert-Nature}, 
with adiabaticity assumed in a number of schemes for geometric quantum
computation (e.g.,~\cite{Pachos:00,Duan-Science:01,Pachos:02,Fazio:03}). Moreover, 
additional interest in adiabatic processes has arisen in connection with 
the concept of adiabatic quantum computing, in
which slowly varying Hamiltonians appear as a promising mechanism for the
design of new quantum algorithms and even as an alternative to the
conventional quantum circuit model of quantum computation~\cite{Farhi:00,Farhi:01}.

Remarkably, the notion of adiabaticity does not appear to have been extended
in a systematic manner to the arena of \emph{open} quantum systems, i.e.,
quantum systems coupled to an external environment \cite{Breuer:book}. 
Such systems are of fundamental interest, as
the notion of a closed system is always an idealization and approximation.
This issue is particularly important in the context of quantum information
processing, where environment-induced decoherence is viewed as a fundamental
obstacle on the path to the construction of quantum information processors
(e.g., \cite{LidarWhaley:03}).

The aim of this work is to systematically
generalize the concept of adiabatic evolution to the realm of open quantum
systems. Formally, an open quantum system is described as follows. Consider
a quantum system $S$ coupled to an environment, or bath $B$ (with respective
Hilbert spaces $\mathcal{H}_{S},\mathcal{H}_{B}$), evolving unitarily under
the total system-bath Hamiltonian $H_{SB}$. The exact system dynamics is
given by tracing over the bath degrees of freedom \cite{Breuer:book} 
\begin{equation}
\rho (t)=\mathrm{Tr}_{B}[U(t)\rho _{SB}(0)U^{\dag }(t)],  \label{system}
\end{equation}%
where $\rho (t)$ is the system state, $\rho _{SB}(0)=\rho (0)\otimes \rho
_{B}(0)$ is the initially uncorrelated system-bath state, and $U(t)=\mathcal{%
T}\mathsf{\exp }[-i\int_{0}^{t}H_{SB}(t^{\prime })dt^{\prime }]$ ($\mathcal{T%
}$ denotes time-ordering; we set $\hbar =1$). Such an evolution is
completely positive and trace preserving
\cite{Breuer:book,Kraus:71,Alicki:87}.
Under certain approximations, it is possible to convert Eq.~(\ref{system})
into the convolutionless form
\begin{eqnarray}
{\dot{\rho}}(t) &=& \mathcal{L}(t) \rho (t).
\label{eq:t-Lind}
\end{eqnarray}
An important example is 
\begin{eqnarray}
{\dot{\rho}}(t) &=& 
-i\left[ H(t),\rho (t) \right] +\frac{1}{2}
\sum_{i=1}^{N}\left([\Gamma _{i}(t),\rho (t) \Gamma^{\dagger }_{i}(t)] \right.
\nonumber \\
&&\left.+[\Gamma_{i}(t)\rho (t), \Gamma^{\dagger }_{i}(t)]\right).
\label{eq:t-Lind2}
\end{eqnarray}%
Here $H(t)$ is the time-dependent effective Hamiltonian of the open system and 
$\Gamma _{i}(t)$ are time-dependent operators describing the system-bath
interaction. In the literature, Eq.~(\ref{eq:t-Lind2}) with time-\emph{in}dependent 
operators $\Gamma _{i}$ is usually referred to as the Markovian dynamical semigroup, 
or Lindblad equation \cite{Breuer:book,Alicki:87,Gorini:76,Lindblad:76} [see also
Ref.~\cite{Lidar:CP01} for a simple derivation of
Eq.~(\ref{eq:t-Lind2}) from Eq.~(\ref{system})]. However,
the case with time-dependent coefficients is also permissible under
certain restrictions \cite{Lendi:86}.
The Lindblad equation requires the
assumption of a Markovian bath with vanishing correlation
time. Equation (\ref{eq:t-Lind}) can be more general; for example, it
applies to the case of non-Markovian convolutionless master equations
studied in Ref.~\cite{Breuer:04}.
In this work we will consider the class of convolutionless master
equations (\ref{eq:t-Lind}). In a slight
abuse of nomenclature, we will henceforth refer to the time-dependent
generator $\mathcal{L}(t)$ as the Lindblad superoperator, and the $\Gamma
_{i}(t)$ as Lindblad operators.

Returning to the problem of adiabatic evolution, conceptually, the
difficulty in the transition from closed to open systems is that the notion
of Hamiltonian eigenstates is lost, since the Lindblad superoperator -- the
generalization of the Hamiltonian -- cannot in general be diagonalized. It
is then not \emph{a priori} clear what should take the place of the adiabatic
eigenstates. Our key insight in resolving this difficulty is that this role
is played by \emph{adiabatic Jordan blocks of the Lindblad superoperator}.
The Jordan canonical form \cite{Horn:book}, with its associated left and
right eigenvectors, is in this context the natural generalization of the
diagonalization of the Hamiltonian. Specifically, we show that, for slowly
varying Lindblad superoperators, the time evolution of the density matrix,
written in a suitable basis in the state space of linear operators, occurs
separately in sets of Jordan blocks related to each Lindblad eigenvalue.
This treatment for adiabatic processes in open systems is potentially rather
attractive as it can simplify the description of the dynamical problem by
breaking down the Lindblad superoperator into a set of decoupled blocks. In
order to clearly exemplify this behavior, we analyze a simple two-level
open system for which the exact solution of the master equation
(\ref{eq:t-Lind}) can be analytically determined.

The paper is organized as follows. We begin, in Sec. \ref{closed}, with a
review of the standard adiabatic approximation for closed quantum systems.
In Sec. \ref{open} we describe the general dynamics of open
quantum systems, review the superoperator formalism, and introduce a
strategy to find suitable bases in the state space of linear operators.
Section \ref{adiabatic} is devoted to deriving our adiabatic approximation,
including the conditions for its validity. In Sec. \ref{example}, we
provide a concrete example which illustrates the consequences of the
adiabatic behavior for systems in the presence of decoherence. Finally, we
present our conclusions in Sec. \ref{conclusions}.

\section{The adiabatic approximation in closed quantum systems}

\label{closed}


\subsection{Condition on the Hamiltonian}

To facilitate comparison with our later derivation of the adiabatic
approximation for open systems, let us begin by reviewing the adiabatic
approximation in closed quantum systems, subject to unitary evolution. In
this case, the evolution is governed by the time-dependent Schr\"{o}dinger
equation 
\begin{equation}
H(t)\,|\psi (t)\rangle =i\,|{\dot{\psi}}(t)\rangle ,  \label{se}
\end{equation}%
where $H(t)$ denotes the Hamiltonian and $|\psi (t)\rangle $ is a quantum
state in a $D$-dimensional Hilbert space. For simplicity, we assume that the
spectrum of $H(t)$ is entirely discrete and nondegenerate. Thus we can
define an instantaneous basis of eigenenergies by 
\begin{equation}
H(t)\,|n(t)\rangle =E_{n}(t)\,|n(t)\rangle ,  \label{ebh}
\end{equation}%
with the set of eigenvectors {$|n(t)\rangle $} chosen to be orthonormal. In
this simplest case, where to each energy level there corresponds a unique
eigenstate, \emph{adiabaticity is then defined as the regime associated to
an independent evolution of the instantaneous eigenvectors of} $H(t)$. This
means that instantaneous eigenstates at one time evolve continuously to the
corresponding eigenstates at later times, and that their corresponding
eigenenergies do not cross. In particular, if the system begins its
evolution in a particular eigenstate $|n(0)\rangle$, then it will evolve to
the instantaneous eigenstate $|n(t)\rangle $ at a later time $t$, without
any transition to other energy levels. In order to obtain a general validity
condition for adiabatic behavior, let us expand $|\psi (t)\rangle $ in terms
of the basis of instantaneous eigenvectors of $H(t)$, 
\begin{equation}
|\psi (t)\rangle =\sum_{n=1}^{D}a_{n}(t)\,e^{-i\int_{0}^{t}dt^{\prime
}E_{n}(t^{\prime })}\,|n(t)\rangle ,  \label{ep}
\end{equation}%
with $a_{n}(t)$ being complex functions of time. Substitution of Eq.~(\ref%
{ep}) into Eq.~(\ref{se}) yields 
\begin{equation}
\sum_{n}\left( {\dot{a}}_{n}|n\rangle +a_{n}|{\dot{n}}\rangle \right)
\,e^{-i\int_{0}^{t}dt^{\prime }E_{n}(t^{\prime })}=0,  \label{an1}
\end{equation}%
where use has been made of Eq.~(\ref{ebh}). Multiplying Eq.~(\ref{an1}) by $%
\langle k(t)|$, we have 
\begin{equation}
{\dot{a}}_{k}=-\sum_{n}a_{n}\langle k|{\dot{n}}\rangle
\,e^{-i\int_{0}^{t}dt^{\prime }g_{nk}(t^{\prime })},  \label{an2}
\end{equation}%
where 
\begin{equation}
g_{nk}(t)\equiv E_{n}(t)-E_{k}(t).  \label{eq:g}
\end{equation}%
A useful expression for $\langle k|{\dot{n}}\rangle $, for $k\neq n$, can be
found by taking the time derivative of Eq.~(\ref{ebh}) and multiplying the
resulting expression by $\langle k|$, which reads 
\begin{equation}
\langle k|{\dot{n}}\rangle =\frac{\langle k|{\dot{H}}|n\rangle }{g_{nk}}%
\quad (n\neq k).  \label{knee}
\end{equation}%
Therefore, Eq.~(\ref{an2}) can be written as 
\begin{equation}
{\dot{a}}_{k}=-a_{k}\langle k|{\dot{k}}\rangle -\sum_{n\neq k}a_{n}\frac{%
\langle k|{\dot{H}}|n\rangle }{g_{nk}}\,e^{-i\int_{0}^{t}dt^{\prime
}g_{nk}(t^{\prime })}.  \label{anf}
\end{equation}%
Adiabatic evolution is ensured if the coefficients $a_{k}(t)$ evolve
independently from each other, i.e., if their dynamical equations do not
couple. As is apparent from Eq.~(\ref{anf}), this requirement is fulfilled
by imposing the conditions 
\begin{equation}
\max_{0\le t\le T} \left\vert \frac{\langle k|{\dot{H}}|n\rangle }{g_{nk}}\right\vert \,\ll\,
\min_{0\le t\le T} \left\vert{g_{nk}}\right\vert,  \label{vcc}
\end{equation}
which serves as an estimate of the validity of the adiabatic
approximation, where $T$ is the total evolution time.   
Note that the left-hand side of Eq.~(\ref{vcc})  
has dimensions of frequency and hence must be compared to the relevant physical frequency scale, 
given by the gap $g_{nk}$~\cite{Messiah:book,Mostafazadeh:book}. 
For a discussion of the adiabatic regime when there is no gap in the energy spectrum see Refs. \cite{Avron:98,Avron:99}.   
In the case of a degenerate spectrum of $H(t)$, Eq.~(\ref{knee}) holds only
for eigenstates $|k\rangle $ and $|n\rangle $ for which $E_{n}\neq E_{k}$.
Taking into account this modification in Eq.~(\ref{anf}), it is not
difficult to see that the adiabatic approximation generalizes to the
statement that each degenerate eigenspace of $H(t)$, instead of individual
eigenvectors, has independent evolution, whose validity conditions given by
Eq.~(\ref{vcc}) are to be considered over eigenvectors with distinct
energies. Thus, in general one can define adiabatic dynamics of closed
quantum systems as follows:

\begin{definition}
\label{defc} A closed quantum system is said to undergo adiabatic dynamics
if its Hilbert space can be decomposed into decoupled Schr\"{o}%
dinger eigenspaces with distinct, time-continuous, and noncrossing
instantaneous eigenvalues of $H(t)$.
\end{definition}

It is conceptually useful to point out that the relationship between slowly
varying Hamiltonians and adiabatic behavior, which explicitly appears from
Eq.~(\ref{vcc}), can also be demonstrated directly from a simple
manipulation of the Schr\"{o}dinger equation: recall that $H(t)$ can be
diagonalized by a unitary similarity tranformation 
\begin{equation}
H_{d}(t)=U^{-1}(t)\,H(t)\,U(t),  \label{hdc}
\end{equation}%
where $H_{d}(t)$ denotes the diagonalized Hamiltonian and $U(t)$ is a
unitary transformation. Multiplying Eq.~(\ref{se}) by $U^{-1}(t)$ and using
Eq.~(\ref{hdc}), we obtain 
\begin{equation}
H_{d}\,|\psi \rangle _{d}=i\,|{\dot{\psi}}\rangle _{d}-i\,{\dot{U}}%
^{-1}|\psi \rangle ,  \label{sed}
\end{equation}%
where $|\psi \rangle _{d}\equiv U^{-1}|\psi \rangle $ is the state
of the system in the basis of eigenvectors of $H(t)$. Upon considering that $%
H(t)$ changes slowly in time, i.e., $dH(t)/dt\approx 0$, we may also assume
that the unitary transformation $U(t)$ and its inverse $U^{-1}(t)$ are
slowly varying operators, yielding 
\begin{equation}
H_{d}(t)\,|\psi (t)\rangle _{d}=i\,|{\dot{\psi}}(t)\rangle _{d}.
\label{eq:Had}
\end{equation}%
Thus, since $H_{d}(t)$ is diagonal, the system evolves separately in each
energy sector, ensuring the validity of the adiabatic approximation. In our
derivation of the condition of adiabatic behavior for open systems below, we
will make use of this semi-intuitive picture in order to motivate the
decomposition of the dynamics into Lindblad-Jordan blocks.

\subsection{Condition on the total evolution time}

The adiabaticity condition can also be given in terms of the total evolution
time $T$. We shall consider for simplicity a nondegenerate $H(t)$; the
generalization to the degenerate case is possible. Let us then rewrite Eq.~(%
\ref{anf}) as follows \cite{Gottfried:book}: 
\begin{equation}
e^{i\gamma _{k}(t)}\,\frac{\partial }{\partial t}[
a_{k}(t)\,e^{-i\gamma _{k}(t)}] = 
-\sum_{n\neq k}a_{n}\frac{\langle k|{\dot{H}}|n\rangle }{g_{nk}}%
\,e^{-i\int_{0}^{t}dt^{\prime }g_{nk}(t^{\prime })},
\label{adtti}
\end{equation}
where $\gamma _{k}(t)$ denotes the Berry's phase \cite{Berry:84} associated
to the state $|k\rangle $, 
\begin{equation}
\gamma _{k}(t)=i\int_{0}^{t}dt^{\prime }\langle k(t^{\prime })|{\dot{k}}%
(t^{\prime })\rangle .
\end{equation}%
Now let us define a normalized time $s$ through the variable transformation 
\begin{equation}
t=sT,\,\,\,\,\,0\leq s\leq 1.  \label{nt}
\end{equation}%
Then, by performing the change $t\rightarrow s$ in Eq.~(\ref{adtti}) and
integrating, we obtain 
\begin{eqnarray}
&&a_{k}(s)\,e^{-i\gamma _{k}(s)}=  \nonumber \\
&&a_{k}(0)-\sum_{n\neq k}\int_{0}^{s}ds^{\prime }\frac{F_{nk}(s^{\prime })}{%
g_{nk}(s^{\prime })}e^{-iT\int_{0}^{s^{\prime }}ds^{\prime \prime
}g_{nk}(s^{\prime \prime })},  \label{akint}
\end{eqnarray}%
where 
\begin{eqnarray}
F_{nk}(s)=a_{n}(s)\,\langle k(s)|\frac{dH(s)}{ds}|n(s)\rangle \,e^{-i\gamma
_{k}(s)}. 
\end{eqnarray}
However, for an adiabatic evolution as defined above, the coefficients $%
a_{n}(s)$ evolve without any mixing, which means that $a_{n}(s)\approx
a_{n}(0)\,e^{i\gamma _{n}(s)}$. Therefore, 
\begin{eqnarray}
F_{nk}(s)=a_{n}(0)\,\langle k(s)|\frac{dH(s)}{ds}|n(s)\rangle \,e^{-i[\gamma
_{k}(s)-\gamma _{n}(s)]}. 
\end{eqnarray}
In order to arrive at a condition on $T$, it is useful to separate out the
fast oscillatory part from Eq.~(\ref{akint}). Thus, the integrand in Eq.~(%
\ref{akint}) can be rewritten as 
\begin{eqnarray}
&&\frac{F_{nk}(s^{\prime })}{g_{nk}(s^{\prime })}%
e^{-iT\int_{0}^{s^{\prime }}ds^{\prime \prime }g_{nk}(s^{\prime \prime })}= 
\nonumber \\
&&\frac{i}{T}\left[ \frac{d}{ds^{\prime }}\left( \frac{%
F_{nk}(s^{\prime })}{g_{nk}^{2}(s^{\prime })}e^{-iT\int_{0}^{s^{\prime
}}ds^{\prime \prime }g_{nk}(s^{\prime \prime })}\right) \right.  \nonumber \\
&&\left. -\,e^{-iT\int_{0}^{s^{\prime }}ds^{\prime \prime
}g_{nk}(s^{\prime \prime })}\frac{d}{ds^{\prime }}\left( \frac{%
F_{nk}(s^{\prime })}{g_{nk}^{2}(s^{\prime })}\right) \right] .  \label{ricc}
\end{eqnarray}%
Substitution of Eq.~(\ref{ricc}) into Eq.~(\ref{akint}) results in 
\begin{eqnarray}
&&a_{k}(s)\,e^{-i\gamma _{k}(s)}=  \nonumber \\
&&a_{k}(0)+\frac{i}{T}\sum_{n\neq k}\left( \frac{F_{nk}(0)}{%
g_{nk}^{2}(0)}-\frac{F_{nk}(s)}{g_{nk}^{2}(s)}e^{-iT\int_{0}^{s}ds^{\prime
}g_{nk}(s^{\prime })}\right.  \nonumber \\
&&\left. +\,\int_{0}^{s}ds^{\prime }\,e^{-iT\int_{0}^{s^{\prime }}ds^{\prime
\prime }g_{nk}(s^{\prime \prime })}\frac{d}{ds^{\prime }}\frac{%
F_{nk}(s^{\prime })}{g_{nk}^{2}(s^{\prime })}\right).  \label{akfinal}
\end{eqnarray}%
A condition for the adiabatic regime can be obtained from Eq.~(\ref{akfinal}%
) if the integral in the last line vanishes for large $T$. Let us assume
that, as $T\rightarrow \infty $, the energy difference remains nonvanishing.
We further assume that $d\{F_{nk}(s^{\prime })/g_{nk}^{2}(s^{\prime
})\}/ds^{\prime }$ is integrable on the interval $\left[ 0,s\right] $. Then
it follows from the Riemann-Lebesgue lemma~\cite{Churchill:book} that the
integral in the last line of Eq.~(\ref{akfinal}) vanishes in the limit $%
T\rightarrow \infty $ (due to the fast oscillation of the integrand)~\cite{RiemannLebesgue}.
What is left are therefore only the first two terms in the sum over $n\neq k$\ of
Eq.~(\ref{akfinal}). Thus, a general estimate of the time rate at which the
adiabatic regime is approached can be expressed by 
\begin{eqnarray}
T\gg \frac{F}{g^{2}}, \label{timead}
\end{eqnarray}
where 
\begin{eqnarray}
&&F=\max_{0\leq s\leq 1}|a_{n}(0)\,\langle k(s)|\frac{dH(s)}{ds}|n(s)\rangle
|,  \nonumber \\
&&g=\min_{0\leq s\leq 1}|g_{nk}(s)|\, ,
\end{eqnarray}%
with $\max $ and $\min $ taken over all $k$ and $n$. A
simplification is obtained if the system starts its evolution in a
particular eigenstate of $H(t)$. Taking the initial state as the eigenvector 
$|m(0)\rangle $, with $a_{m}(0)=1$, adiabatic evolution occurs if 
\begin{eqnarray}
T\gg \frac{\mathcal{F}}{\mathcal{G}^{2}}, \label{timead2}
\end{eqnarray}
where 
\begin{eqnarray}
&&\mathcal{F}=\max_{0\leq s\leq 1}|\langle k(s)|\frac{dH(s)}{ds}|m(s)\rangle
|\,,  \nonumber \\
&&\mathcal{G}=\min_{0\leq s\leq 1}|g_{mk}(s)|\,.
\end{eqnarray}%
Equation~(\ref{timead2}) gives an important validity condition for the adiabatic
approximation, which has been used, e.g., to determine the running time
required by adiabatic quantum algorithms~\cite{Farhi:00,Farhi:01}.


\section{The dynamics of open quantum systems}

\label{open}


In this section, we prepare the mathematical framework required to derive an
adiabatic approximation for open quantum systems. Our starting point is the
convolutionless master equation (\ref{eq:t-Lind}). It proves
convenient to transform to the superoperator formalism, wherein the density
matrix is represented by a $D^{2}$-dimensional \textquotedblleft coherence
vector\textquotedblright\ 
\begin{equation}
|\rho \rangle \rangle =\left( 
\begin{array}{cccc}
\rho _{1} & \rho _{2} & \cdots  & \rho _{D^{2}}%
\end{array}%
\right) ^{t},  \label{vcv}
\end{equation}%
and the Lindblad superoperator $\mathcal{L}$ becomes a 
$(D^{2}\times D^{2})$-dimensional supermatrix \cite{Alicki:87}. We use the double bracket
notation to indicate that we are not working in the standard Hilbert space
of state vectors. Such a representation can be generated, e.g., by
introducing a basis of Hermitian, trace-orthogonal, and traceless operators
[e.g., su($D$)], whence the $\rho _{i}$ are the expansion coefficients of $%
\rho $ in this basis \cite{Alicki:87}, with $\rho _{1}$ the coefficient of 
$I$ (the identity matrix). In this case, the condition $\mathrm{Tr}\rho ^{2}\leq 1$ corresponds to $%
\left\Vert |\rho \rangle \rangle \right\Vert \leq 1$, $\rho =\rho ^{\dag }$
to $\rho _{i}=\rho _{i}^{\ast }$, and positive semidefiniteness of $\rho $
is expressed in terms of inequalities satisfied by certain Casimir
invariants [e.g., of $su(D)$] \cite{byrd:062322,Kimura:2003-1,Kimura:2003-2}. A\ simple and well-known
example of this procedure is the representation of the density operator of a
two-level system (qubit) on the Bloch sphere, via $\rho =(I_{2}+%
\overrightarrow{v}\cdot \overrightarrow{\sigma })/2$, where $\overrightarrow{%
\sigma }=(\sigma _{x},\sigma _{y},\sigma _{z})$ is the vector of Pauli
matrices, $I_{2}$ is the $2\times 2$ identity matrix, and $\overrightarrow{v}%
\in \mathbb{R}^{3}$ is a three-dimensional coherence vector of norm$\leq 1$.
More generally, coherence vectors live in Hilbert-Schmidt space:\ a state
space of linear operators endowed with an inner product that can be defined,
for general vectors $u$ and $v$, as 
\begin{equation}
(u,v)\equiv \langle \langle u|v\rangle \rangle \equiv \frac{1}{{\mathcal{N}}}{%
\text{Tr}}\left( u^{\dagger }v\right) ,  \label{ip}
\end{equation}%
where ${\mathcal{N}}$ is a normalization factor. Adjoint elements $\langle
\langle v|$ in the dual state space are given by row vectors defined as the
transpose conjugate of $|v\rangle \rangle $: $\langle \langle
v|=(v_{1}^{\ast },v_{2}^{\ast },...,v_{D^{2}}^{\ast })$. A density matrix
can then be expressed as a discrete superposition of states over a complete
basis in this vector space, with appropriate constraints on the coefficients
so that the requirements of Hermiticity, positive semidefiniteness, and unit
trace of $\rho $ are observed. Thus, representing the density operator in
general as a coherence vector, we can rewrite Eq.~(\ref{eq:t-Lind}) in a
superoperator language as
\begin{equation}
\mathcal{L}(t)\,|\rho (t)\rangle \rangle =|{\dot{\rho}}(t)\rangle \rangle ,
\label{le}
\end{equation}%
where $\mathcal{L}$ is now a supermatrix. This master equation generates
nonunitary evolution, since $\mathcal{L}(t)$ is non-Hermitian and hence
generally nondiagonalizable. However, it is always possible to obtain an
elegant decomposition in terms of a block structure, the Jordan canonical
form \cite{Horn:book}. This can be achieved by the similarity
transformation 
\begin{equation}
\mathcal{L}_{J}(t)=S^{-1}(t)\,\mathcal{L}(t)\,S(t),  \label{jd}
\end{equation}%
where $\mathcal{L}_{J}(t)=\mathrm{diag}(J_{1},...,J_{m})$ denotes the Jordan
form of $\mathcal{L}(t)$, with $J_{\alpha }$ representing a Jordan block
related to an eigenvector whose corresponding eigenvalue is $\lambda
_{\alpha }$, 
\begin{equation}
J_{\alpha }=\left( 
\begin{array}{ccccc}
\lambda _{\alpha } & 1 & 0 & \cdots & 0 \\ 
0 & \lambda _{\alpha } & 1 & \cdots & 0 \\ 
\vdots  & \ddots  & \ddots  & \ddots  & \vdots  \\ 
0 & \cdots & 0 & \lambda _{\alpha } & 1 \\ 
0 & \cdots & \cdots & 0 & \lambda _{\alpha }%
\end{array}%
\right) .  \label{ljmatg}
\end{equation}%
The number $m$ of Jordan blocks is given by the number of linearly
independent eigenstates of $\mathcal{L}(t)$, with each eigenstate associated
to a different block $J_{\alpha }$. Since $\mathcal{L}(t)$ is in general non-Hermitian,
we generally do not have a basis of eigenstates, whence some care is
required in order to find a basis for describing the density operator. A
systematic procedure for finding a convenient discrete vector basis is to
start from the instantaneous right and left eigenstates of $\mathcal{L}(t)$,
which are defined by 
\begin{eqnarray}
\mathcal{L}(t)\,|\mathcal{P}_{\alpha }(t)\rangle \rangle  &=&\lambda
_{\alpha }(t)\,|\mathcal{P}_{\alpha }(t)\rangle \rangle ,  \label{rleb0} \\
\langle \langle Q_{\alpha }(t)|\,\mathcal{L}(t) &=&\langle \langle Q_{\alpha
}(t)|\,\lambda _{\alpha }(t),  \label{rleb}
\end{eqnarray}%
where, in our notation, possible degeneracies correspond to $\lambda
_{\alpha }=\lambda _{\beta }$, with $\alpha \neq \beta $. In other words, we
reserve a different index $\alpha $ for each independent eigenvector since
each eigenvector is in a distinct Jordan block. It can immediately be shown
from Eqs.~(\ref{rleb0}) and~(\ref{rleb}) that, for $\lambda _{\alpha }\neq
\lambda _{\beta }$, we have $\langle \langle Q_{\alpha }(t)|\mathcal{P}%
_{\beta }(t)\rangle \rangle =0$. The left and right eigenstates can be
easily identified when the Lindblad superoperator is in the Jordan form $%
\mathcal{L}_{J}(t)$. Denoting $|\mathcal{P}_{\alpha }(t)\rangle \rangle
_{J}=S^{-1}(t)\,|\mathcal{P}_{\alpha }(t)\rangle \rangle $, i.e., the right
eigenstate of $\mathcal{L}_{J}(t)$ associated to a Jordan block $J_{\alpha }$%
, then Eq.~(\ref{rleb0}) implies that $|\mathcal{P}_{\alpha }(t)\rangle
\rangle _{J}$ is time-independent and, after normalization, is given by 
\begin{equation}
\left. |\mathcal{P}_{\alpha }\rangle \rangle _{J}\frac{{}}{{}}\right\vert
_{J_{\alpha }}=\left( 
\begin{array}{c}
1 \\ 
0 \\ 
\vdots  \\ 
0 \\ 
\end{array}%
\right) ,  \label{pj}
\end{equation}%
where only the vector components associated to the Jordan block $J_{\alpha }$
are shown, with all the others vanishing. In order to have a complete basis
we shall define new states, which will be chosen so that they preserve the
block structure of $\mathcal{L}_{J}(t)$. A suitable set of additional
vectors is 
\begin{equation}
\left. |\mathcal{D}_{\alpha }^{(1)}\rangle \rangle _{J}\frac{{}}{{}}%
\right\vert _{J_{\alpha }}=\left( 
\begin{array}{c}
0 \\ 
1 \\ 
0 \\ 
\vdots  \\ 
0 \\ 
\end{array}%
\right) ,\,...\,,\,\left. |\mathcal{D}_{\alpha }^{(n_{\alpha }-1)}\rangle
\rangle _{J}\frac{{}}{{}}\right\vert _{J_{\alpha }}=\left( 
\begin{array}{c}
0 \\ 
0 \\ 
0 \\ 
\vdots  \\ 
1 \\ 
\end{array}%
\right) ,  \label{dj}
\end{equation}%
where $n_{\alpha }$ is the dimension of the Jordan block $J_{\alpha }$ and
again all the components outside $J_{\alpha }$ are zero. This simple vector
structure allows for the derivation of the expression 
\begin{equation}
\mathcal{L}_{J}(t)\,|\mathcal{D}_{\alpha }^{(j)}\rangle \rangle _{J}=|%
\mathcal{D}_{\alpha }^{(j-1)}\rangle \rangle _{J}+\lambda _{\alpha }(t)\,|%
\mathcal{D}_{\alpha }^{(j)}\rangle \rangle _{J},  \label{ldj}
\end{equation}%
with $|\mathcal{D}_{\alpha }^{(0)}\rangle \rangle _{J}\equiv |\mathcal{P}%
_{\alpha }\rangle \rangle _{J}$ and \ $|\mathcal{D}_{\alpha }^{(-1)}\rangle
\rangle _{J}\equiv 0$. The set $\left\{ |\mathcal{D}_{\alpha }^{(j)}\rangle
\rangle _{J},\text{\thinspace with}\,j=0,...,(n_{\alpha }-1)\right\} $ can
immediately be related to a right vector basis for the original $\mathcal{L}%
(t)$ by means of the transformation $|\mathcal{D}_{\alpha }^{(j)}(t)\rangle
\rangle =S(t)\,|\mathcal{D}_{\alpha }^{(j)}\rangle \rangle _{J}$ which,
applied to Eq.~(\ref{ldj}), yields 
\begin{equation}
\mathcal{L}(t)\,|\mathcal{D}_{\alpha }^{(j)}(t)\rangle \rangle =|\mathcal{D}%
_{\alpha }^{(j-1)}(t)\rangle \rangle +\lambda _{\alpha }(t)\,|\mathcal{D}%
_{\alpha }^{(j)}(t)\rangle \rangle .  \label{ldo}
\end{equation}%
Equation~(\ref{ldo}) exhibits an important feature of the set $\left\{ |\mathcal{D%
}_{\beta }^{(j)}(t)\rangle \rangle \right\} $, namely, it implies that
Jordan blocks are invariant under the action of the Lindblad superoperator.
An analogous procedure can be employed to define the left eigenbasis.
Denoting by $_{J}\langle \langle \mathcal{Q}_{\alpha }(t)|=\langle \langle 
\mathcal{Q}_{\alpha }(t)|S(t)$ the left eigenstate of $\mathcal{L}_{J}(t)$
associated to a Jordan block $J_{\alpha }$, Eq.~(\ref{rleb}) leads to the
normalized left vector 
\begin{equation}
\left. _{J}\langle \langle \mathcal{Q}_{\alpha }|\frac{{}}{{}}\right\vert
_{J_{\alpha }}=\left( \frac{{}}{{}}0,\,.\,.\,.\,,0,1\frac{{}}{{}}\right) .
\label{qj}
\end{equation}%
The additional left vectors are defined as 
\begin{eqnarray}
\left. _{J}\langle \langle \mathcal{E}_{\alpha }^{(0)}|\frac{{}}{{}}%
\right\vert _{J_{\alpha }} &=&\left( \frac{{}}{{}}1,0,0,\,.\,.\,.\,,0\frac{{}%
}{{}}\right) ,  \nonumber \\
\vspace{0.1cm} &.\,\,.\,\,.&  \nonumber \\
\vspace{0.1cm}\left. _{J}\langle \langle \mathcal{E}_{\alpha }^{(n_{\alpha
}-2)}|\frac{{}}{{}}\right\vert _{J_{\alpha }} &=&\left( \frac{{}}{{}}%
0,\,.\,.\,.\,,0,1,0\frac{{}}{{}}\right) ,  \label{rj}
\end{eqnarray}%
which imply the following expression for the left basis vector $\langle
\langle \mathcal{E}_{\alpha }^{(i)}(t)|=\,_{J}\langle \langle \mathcal{E}%
_{\alpha }^{(i)}|\,S^{-1}(t)$ for $\mathcal{L}(t)$: 
\begin{equation}
\langle \langle \mathcal{E}_{\alpha }^{(i)}(t)|\,\mathcal{L}(t)=\langle
\langle \mathcal{E}_{\alpha }^{(i+1)}(t)|+\langle \langle \mathcal{E}%
_{\alpha }^{(i)}(t)|\,\lambda _{\alpha }(t).  \label{lro}
\end{equation}%
Here we have used the notation $_{J}\langle \langle \mathcal{E}_{\alpha
}^{(n_{\alpha }-1)}|\equiv \,_{J}\langle \langle \mathcal{Q}_{\alpha }|$ and 
$_{J}\langle \langle \mathcal{E}_{\alpha }^{(n_{\alpha })}|\equiv 0$. A
further property following from the definition of the right and left vector
bases introduced here is 
\begin{equation}
\langle \langle \mathcal{E}_{\alpha }^{(i)}(t)|\mathcal{D}_{\beta
}^{(j)}(t)\rangle \rangle =\,_{J}\langle \langle \mathcal{E}_{\alpha }^{(i)}|%
\mathcal{D}_{\beta }^{(j)}\rangle \rangle _{J}=\delta _{\alpha \beta }\delta
^{ij}.  \label{lrr}
\end{equation}%
This orthonormality relationship between corresponding left and right states
will be very useful in our derivation below of the conditions for the
validity of the adiabatic approximation.

\section{The adiabatic approximation in open quantum systems}

\label{adiabatic}


We are now ready to derive our main result:\ an adiabatic approximation for
open quantum systems. We do this by observing that the Jordan decomposition
of $\mathcal{L}(t)$ [Eq.~(\ref{jd})] allows for a nice
generalization of the standard quantum adiabatic approximation. We begin by
defining the adiabatic dynamics of an open system as a generalization of the
definition given above for closed quantum systems:

\begin{definition}
\label{def:open-ad} An open quantum system is said to undergo adiabatic
dynamics if its Hilbert-Schmidt space can be decomposed into decoupled
Lindblad--Jordan eigenspaces with distinct, time-continuous, and
noncrossing instantaneous eigenvalues of $\mathcal{L}(t)$.
\end{definition}

This definition is a natural extension for open systems of the idea of
adiabatic behavior. Indeed, in this case the master equation (\ref{eq:t-Lind}%
) can be decomposed into sectors with different and separately evolving
Lindblad-Jordan eigenvalues, and we show below that the condition for this
to occur is appropriate \textquotedblleft slowness\textquotedblright\ of the
Lindblad superoperator. The splitting into Jordan blocks of the Lindblad
superoperator is achieved through the choice of a basis which preserves the
Jordan block structure as, for example, the sets of right $\left\{ |\mathcal{%
D}_{\beta }^{(j)}(t)\rangle \rangle \right\} $ and left $\left\{ \langle
\langle \mathcal{E}_{\alpha }^{(i)}(t)|\right\} $ vectors introduced in
Sec. \ref{open}. Such a basis generalizes the notion of Schr\"{o}%
dinger eigenvectors.

\subsection{Intuitive derivation}

Let us first show how the adiabatic Lindblad-Jordan blocks arise from a
simple argument, analogous to the one presented for the closed case [Eqs.~(%
\ref{hdc})-(\ref{eq:Had})]. Multiplying Eq.~(\ref{le}) by the similarity
transformation matrix $S^{-1}(t)$, we obtain 
\begin{eqnarray}
\mathcal{L}_{J}\,|\rho \rangle \rangle _{J}=|{\dot{\rho}}\rangle \rangle
_{J}-{\dot{S}}^{-1}\,|\rho \rangle \rangle , \label{ljinter}
\end{eqnarray}
where we have used Eq.~(\ref{jd}) and defined $|\rho \rangle \rangle
_{J}\equiv S^{-1}|\rho \rangle \rangle $. Now suppose that $\mathcal{L}(t)$,
and consequently $S(t)$ and its inverse $S^{-1}(t)$, changes slowly in time
so that ${\dot{S}}^{-1}(t)\approx 0$. Then, from Eq.~(\ref{ljinter}), the
adiabatic dynamics of the system reads 
\begin{equation}
\mathcal{L}_{J}(t)\,|\rho (t)\rangle \rangle _{J}=|{\dot{\rho}}(t)\rangle
\rangle _{J}.  \label{ljrj}
\end{equation}%
Equation~(\ref{ljrj}) ensures that, choosing an instantaneous basis for the
density operator $\rho (t)$ which preserves the Jordan block structure, the
evolution of $\rho (t)$ occurs separately in adiabatic blocks associated
with distinct eigenvalues of $\mathcal{L}(t)$. Of course, the conditions
under which the approximation ${\dot{S}}^{-1}(t)\approx 0$ holds must be
carefully clarified. This is the subject of the next two subsections.

\subsection{Condition on the Lindblad superoperator}

Let us now derive the validity conditions for open-system adiabatic dynamics
by analyzing the general time evolution of a density operator under the
master equation~(\ref{le}). To this end, we expand the density matrix for an
arbitrary time $t$ in the instantaneous right eigenbasis $\left\{ |{\mathcal{%
D}_{\beta }^{(j)}(t)\rangle \rangle }\right\} $ as 
\begin{equation}
|\rho (t)\rangle \rangle =\frac{1}{2}\sum_{\beta =1}^{m}\sum_{j=0}^{n_{\beta
}-1}r_{\beta }^{(j)}(t)\,|\mathcal{D}_{\beta }^{(j)}(t)\rangle \rangle ,
\label{rtime}
\end{equation}%
where $m$ is the number of Jordan blocks and $n_{\beta }$ is the dimension
of the block $J_{\beta }$. We emphasize that we are assuming that there are
no eigenvalue crossings in the spectrum of the Lindblad superoperator during
the evolution. Requiring then that the density operator Eq.~(\ref{rtime})
evolves under the master equation~(\ref{le}) and making use of Eq.~(\ref{ldo}), we obtain 
\begin{eqnarray}
\sum_{\beta =1}^{m}\sum_{j=1}^{n_{\beta }-1}r_{\beta }^{(j)}\,\left( |%
\mathcal{D}_{\beta }^{(j-1)}\rangle \rangle +\lambda _{\beta }\,|\mathcal{D}%
_{\beta }^{(j)}\rangle \rangle \right) =  \nonumber \\
\sum_{\beta =1}^{m}\sum_{j=0}^{n_{\beta }-1}\left( {\dot{r}}_{\beta
}^{(j)}\,|\mathcal{D}_{\beta }^{(j)}\rangle \rangle +r_{\beta }^{(j)}\,|{%
\dot{\mathcal{D}}}_{\beta }^{(j)}\rangle \rangle \right) .  \label{lindg1}
\end{eqnarray}%
Equation~(\ref{lindg1}) multiplied by the left eigenstate $\langle \langle 
\mathcal{E}_{\alpha }^{(i)}|$ results in 
\begin{equation}
{\dot{r}}_{\alpha }^{(i)}=\lambda _{\alpha }\,r_{\alpha }^{(i)}+r_{\alpha
}^{(i+1)}-\sum_{\beta =1}^{m}\sum_{j=0}^{n_{\beta }-1}r_{\beta
}^{(j)}\langle \langle \mathcal{E}_{\alpha }^{(i)}|{\dot{\mathcal{D}}}%
_{\beta }^{(j)}\rangle \rangle ,  \label{rdot}
\end{equation}%
with $r_{\alpha }^{(n_{\alpha })}(t)\equiv 0$. Note that the sum over $\beta 
$ mixes different Jordan blocks. An analogous situation occurred in the
closed system case, in Eq.~(\ref{anf}). Similarly to what was done there, in
order to derive an adiabaticity condition we must separate this sum into
terms related to the eigenvalue $\lambda _{\alpha }$ of $\mathcal{L}(t)$ and
terms involving mixing with eigenvalues $\lambda _{\beta }\neq \lambda
_{\alpha }$. In this latter case, an expression can be found for $\langle
\langle \mathcal{E}_{\alpha }^{(i)}|{\dot{\mathcal{D}}}_{\beta
}^{(j)}\rangle \rangle $ as follows: taking the time derivative of Eq.~(\ref%
{ldo}) and multiplying by $\langle \langle \mathcal{E}_{\alpha }^{(i)}|$ we
obtain, after using Eqs.~(\ref{lro}) and~(\ref{lrr}), 
\begin{eqnarray}
\langle \langle \mathcal{E}_{\alpha }^{(i)}|{\dot{\mathcal{D}}}_{\beta
}^{(j)}\rangle \rangle =\frac{1}{\omega _{\beta \alpha }}\,\left(
\,\langle \langle \mathcal{E}_{\alpha }^{(i)}|\,{\dot{\mathcal{L}}}\,|%
\mathcal{D}_{\beta }^{(j)}\rangle \rangle \,\right. \nonumber
\\
\left. +\,\langle \langle \mathcal{E}_{\alpha }^{(i+1)}|{\dot{\mathcal{D}}}%
_{\beta }^{(j)}\rangle \rangle -\langle \langle \mathcal{E}_{\alpha }^{(i)}|{%
\dot{\mathcal{D}}}_{\beta }^{(j-1)}\rangle \rangle \,\right) ,\,\,\,\,
\label{rddi}
\end{eqnarray}%
where we have defined 
\begin{equation}
\omega _{\beta \alpha }(t)\equiv \lambda _{\beta }(t)-\lambda _{\alpha }(t)
\label{eq:omab}
\end{equation}%
and assumed $\lambda _{\alpha }\neq \lambda _{\beta }$. Note that, while $%
\omega _{\beta \alpha }$ plays a role analogous to that of the energy
difference $g_{nk}$ in the closed case [Eq.~(\ref{eq:g})], $\omega _{\beta
\alpha }$ may be complex. A similar procedure can generate expressions for
all the terms $\langle \langle \mathcal{E}_{\alpha }^{(i)}|{\dot{\mathcal{D}}%
}_{\beta }^{(j-k)}\rangle \rangle $, with $k=0,...,j$. Thus, an iteration of
Eq.~(\ref{rddi}) yields 
\begin{eqnarray}
\langle \langle \mathcal{E}_{\alpha }^{(i)}|{\dot{\mathcal{D}}}_{\beta
}^{(j)}\rangle \rangle &=&\sum_{k=0}^{j}\frac{(-1)^{k}}{\omega _{\beta
\alpha }^{k+1}}\left( \langle \langle \mathcal{E}_{\alpha }^{(i)}|\,{\dot{%
\mathcal{L}}}\,|\mathcal{D}_{\beta }^{(j-k)}\rangle \rangle \,\right. 
 \nonumber \\
&&\left. +\,\langle \langle \mathcal{E}_{\alpha }^{(i+1)}|{\dot{\mathcal{D}}}%
_{\beta }^{(j-k)}\rangle \rangle \right) .  \label{rddi3}
\end{eqnarray}
From a second recursive iteration, now for the term $\langle \langle 
\mathcal{E}_{\alpha }^{(i+1)}|{\dot{\mathcal{D}}}_{\beta }^{(j-k)}\rangle
\rangle $ in Eq.~(\ref{rddi3}), we obtain 
\begin{eqnarray}
&& \langle \langle \mathcal{E}_{\alpha }^{(i)}|{\dot{\mathcal{D}}}_{\beta
}^{(j)}\rangle \rangle =  \nonumber \\
&& \sum_{p=1}^{(n_{\alpha }-i)}\left(
\prod_{q=1}^{p}\sum_{k_{q}=0}^{(j-S_{q-1})}\right) \frac{\langle \langle 
\mathcal{E}_{\alpha }^{(i+p-1)}|{\dot{\mathcal{L}}}|\mathcal{D}_{\beta
}^{(j-S_{p})}\rangle \rangle }{(-1)^{S_{p}}\,\omega _{\beta \alpha
  }^{p+S_{p}}}, \label{rdd}
\end{eqnarray}%
where 
\begin{equation}
S_{q}=\sum_{s=1}^{q}k_{s}\,\,\,\,,\,\,\,\,\left(
\prod_{q=1}^{p}\sum_{k_{q}=0}^{(j-S_{q-1})}\right)
=\sum_{k_{1}=0}^{j-S_{0}}\cdots \sum_{k_{p}=0}^{j-S_{p-1}},
\end{equation}%
with $S_{0}=0$. We can now split Eq.~(\ref{rdot}) into diagonal and
off-diagonal terms 
\begin{eqnarray}
&&{\dot{r}}_{\alpha }^{(i)}=\lambda _{\alpha }r_{\alpha
}^{(i)}+r_{\alpha }^{(i+1)}-\sum_{\beta \,|\,\lambda _{\beta }=\lambda
_{\alpha }}\sum_{j=0}^{n_{\beta }-1}r_{\beta }^{(j)}\langle \langle \mathcal{%
E}_{\alpha }^{(i)}|{\dot{\mathcal{D}}}_{\beta }^{(j)}\rangle \rangle \hspace{%
0.3cm}  \nonumber \\
&&-\sum_{\beta \,|\,\lambda _{\beta }\neq \lambda _{\alpha
}}\sum_{j=0}^{n_{\beta }-1}r_{\beta }^{(j)}\langle \langle \mathcal{E}%
_{\alpha }^{(i)}|{\dot{\mathcal{D}}}_{\beta }^{(j)}\rangle \rangle ,
\label{rfinal}
\end{eqnarray}%
where the terms $\langle \langle \mathcal{E}_{\alpha }^{(i)}|{\dot{\mathcal{D%
}}}_{\beta }^{(j)}\rangle \rangle $, for $\lambda _{\beta }\neq \lambda
_{\alpha }$, are given by Eq.~(\ref{rdd}). In accordance with our definition
of adiabaticity above, the adiabatic regime is obtained when the sum in the
second line is negligible. Summarizing, by introducing the normalized time $s$ defined by 
Eq.~(\ref{nt}), we thus find the following from Eqs.~(\ref{rdd}) and~(\ref{rfinal}).

\begin{theorem}
  \label{t1}
  A sufficient condition for open quantum system adiabatic dynamics as
given in Definition~\ref{def:open-ad} is: 
\begin{eqnarray}
&&\max_{0\le s\le 1}\,\left\vert \sum_{p=1}^{(n_{\alpha }-i)}\left(
\prod_{q=1}^{p}\sum_{k_{q}=0}^{(j-S_{q-1})}\right) \frac{\langle \langle 
\mathcal{E}_{\alpha }^{(i+p-1)}|\frac{d\mathcal{L}}{ds}|\mathcal{D}_{\beta
}^{(j-S_{p})}\rangle \rangle }{(-1)^{S_{p}}\,\omega _{\beta \alpha
  }^{p+S_{p}}}\right\vert \nonumber \\
&&\ll 1, 
\label{vc}
\end{eqnarray}%
with $\lambda _{\beta }\neq \lambda _{\alpha }$ and for arbitrary indices $i$
and $j$ associated to the Jordan blocks $\alpha $ and $\beta $, respectively.
\end{theorem}

The condition (\ref{vc}) ensures the absence of mixing of coefficients $r_{\alpha
}^{(i)}$ related to distinct eigenvalues $\lambda _{\alpha }$ in Eq.~(\ref%
{rfinal}), which in turn guarantees that sets of Jordan blocks belonging to
different eigenvalues of $\mathcal{L}(t)$ have independent evolution. Thus
the accuracy of the adiabatic approximation can be estimated by the
computation of the time derivative of the Lindblad superoperator acting on
right and left vectors. Equation~(\ref{vc}) can be simplified by considering the 
term with maximum absolute value, which results in:

\begin{corollary}
\label{c1os}
A sufficient condition for open quantum system adiabatic dynamics is 
\begin{eqnarray}
&&{\cal N}_{ij}^{n_\alpha n_\beta} \max_{0\le s\le 1}\,\left\vert \frac{\langle \langle 
\mathcal{E}_{\alpha }^{(i+p-1)}|\frac{d\mathcal{L}}{ds}|\mathcal{D}_{\beta
}^{(j-S_{p})}\rangle \rangle }{\omega _{\beta \alpha
  }^{p+S_{p}}}\right\vert \ll 1, 
\end{eqnarray}%
where the $\max$ is taken for any $\alpha \ne \beta$, and over all possible values 
of $i\in\{0,...,n_{\alpha}-1\}$, $j\in\{0,...,n_{\beta}-1\}$, and $p$, with 
\begin{eqnarray}
&&{\cal N}_{ij}^{n_\alpha n_\beta} = \sum_{p=1}^{(n_{\alpha }-i)}\left(
\prod_{q=1}^{p}\sum_{k_{q}=0}^{(j-S_{q-1})}\right) 1 
\label{numberterms} \\
&&= \left(
\begin{array}
[c]{c}
n_\alpha-i+1+j\\
1+j
\end{array}
\right)-1 
= \frac{(n_\alpha-i+1+j)!}{(1+j)!(n_\alpha-i)!}-1. \nonumber
\end{eqnarray}
\end{corollary}
Observe that the factor ${\cal N}_{ij}^{n_\alpha n_\beta}$ defined in Eq.~(\ref{numberterms}) is just the 
number of terms of the sums in Eq.~(\ref{vc}). We have included a
superscript $n_\beta$, even though there is no explicit dependence on
$n_\beta$, since $j\in\{0,...,n_{\beta}-1\}$. 

Furthermore, an adiabatic condition for a slowly varying Lindblad super-operator can 
directly be obtained from Eq.~(\ref{vc}), yielding the following.

\begin{corollary}
A simple sufficient condition for open quantum system adiabatic dynamics is ${%
\dot{\mathcal{L}}}\approx 0$.
\end{corollary}

Note that this condition is in a sense too strong, since it need not
be the case that $\dot{\mathcal{L}}$ is small in general (i.e., for all its matrix elements). 
Indeed, in Sec.~\ref{example} we show via
an example that adiabaticity may occur due to the \emph{exact}
vanishing of relevant matrix elements of ${\dot{\mathcal{L}}}$. The general condition for this 
to occur is the presence of a \emph{dynamical symmetry} \cite{Bohm:88}.

Let us end this subsection by mentioning that we can also write Eq.~(\ref{vc}) in terms of 
the time variable $t$ instead of the normalized time $s$. In this case, the natural 
generalization of Eq.~(\ref{vc}) is 
\begin{eqnarray}
&&\max_{0\le t\le T}\,\left\vert \sum_{p=1}^{(n_{\alpha }-i)}\left(
\prod_{q=1}^{p}\sum_{k_{q}=0}^{(j-S_{q-1})}\right) \frac{\langle \langle 
\mathcal{E}_{\alpha }^{(i+p-1)}|{\dot{\mathcal{L}}}|\mathcal{D}_{\beta
}^{(j-S_{p})}\rangle \rangle }{(-1)^{S_{p}}\,\omega _{\beta \alpha
  }^{p+S_{p}}}\right\vert  \nonumber \\
&&\ll \,\min_{0\le t\le T} \, |\omega_{\beta \alpha}|. 
\label{vch}
\end{eqnarray}
Note that, as in the analogous condition (\ref{vcc}) in the closed case,
the left-hand side has dimensions of frequency, and hence must be
compared to the natural frequency scale $\omega_{\beta \alpha}$. However, 
unlike the closed systems case, where Eq.~(\ref{vcc}) can 
immediately be derived from the time condition (\ref{timead}), we cannot prove here that $\omega_{\beta\alpha}$ 
is indeed the relevant physical scale. 
Therefore, Eq.~(\ref{vch}) should be regarded as a heuristic criterion.

\subsection{Condition on the total evolution time}

\label{sec:tot-t}

As mentioned in Sec.~\ref{closed}, for closed systems the rate at which
the adiabatic regime is approached can be estimated in terms of the total
time of evolution, as shown by Eqs.~(\ref{timead}) and~(\ref{timead2}). We
now provide a generalization of this estimate for adiabaticity in open
systems.

\subsubsection{One-dimensional Jordan blocks}

Let us begin by considering the particular case where $\mathcal{L}(t)$ has
only one-dimensional Jordan blocks and each eigenvalue corresponds to a
single independent eigenvector, i.e., $\lambda _{\alpha }=\lambda _{\beta
}\Rightarrow \alpha =\beta $. Bearing these assumptions in mind, Eq.~(\ref%
{rfinal}) can be rewritten as 
\begin{equation}
{\dot{r}}_{\alpha }=\lambda _{\alpha }r_{\alpha }-r_{\alpha }\langle \langle 
\mathcal{E}_{\alpha }|{\dot{\mathcal{D}}}_{\alpha }\rangle \rangle
-\sum_{\beta \neq \alpha }r_{\beta }\langle \langle \mathcal{E}_{\alpha }|{%
\dot{\mathcal{D}}}_{\beta }\rangle \rangle ,  \label{rfsc}
\end{equation}%
where the upper indices $i,j$ have been removed since we are considering
only one-dimensional blocks. Moreover, for this special case, we have from
Eq.~(\ref{rdd}) 
\begin{equation}
\langle \langle \mathcal{E}_{\alpha }|{\dot{\mathcal{D}}}_{\beta }\rangle
\rangle =\frac{\langle \langle \mathcal{E}_{\alpha }|\,{\dot{\mathcal{L}}}\,|%
\mathcal{D}_{\beta }\rangle \rangle }{\omega _{\beta \alpha }}.  \label{edsc}
\end{equation}%
In order to eliminate the term $\lambda _{\alpha }r_{\alpha }$ from Eq.~(\ref%
{rfsc}), we redefine the variable $r_{\alpha }(t)$ as 
\begin{equation}
r_{\alpha }(t)=p_{\alpha }(t)\,e^{\int_{0}^{t}\lambda _{\alpha }(t^{\prime
})dt^{\prime }},
\end{equation}%
which, applied to Eq.~(\ref{rfsc}), yields 
\begin{equation}
{\dot{p}}_{\alpha }=-p_{\alpha }\,\langle \langle \mathcal{E}%
_{\alpha }|{\dot{\mathcal{D}}}_{\alpha }\rangle \rangle -\sum_{\beta \neq
\alpha }p_{\beta }\,\langle \langle \mathcal{E}_{\alpha }|{\dot{\mathcal{D}}}%
_{\beta }\rangle \rangle \,e^{\Omega _{\beta \alpha }},  \label{eq:rfsc2}
\end{equation}%
with 
\begin{equation}
\Omega _{\beta \alpha }(t)=\int_{0}^{t}dt^{\prime }\,\omega _{\beta \alpha
}(t^{\prime }).  \label{omos}
\end{equation}%
Equation~(\ref{eq:rfsc2}) is very similar to Eq.~(\ref{anf}) for closed systems,
but the fact that $\Omega _{\beta \alpha }$ is in general complex-valued
leads to some important differences, discussed below. We next introduce the
scaled time $s=t/T$ and integrate the resulting expression. Using Eq.~(\ref%
{edsc}), we then obtain 
\begin{eqnarray}
&&p_{\alpha }(s)\,=\,p_{\alpha }(0)-\int_{0}^{s}ds^{\prime
}p_{\alpha }(s^{\prime })\,\Phi _{\alpha }(s^{\prime })  \nonumber \\
&&-\sum_{\beta \neq \alpha }\int_{0}^{s}ds^{\prime }\frac{%
V_{\beta \alpha }(s^{\prime })}{\omega _{\beta \alpha }(s^{\prime })}%
\,e^{T\,\Omega _{\beta \alpha }(s^{\prime })},  \label{scint}
\end{eqnarray}%
where $\Phi _{\alpha }(s)$ is defined by 
\begin{equation}
\Phi _{\alpha }(s)=\langle \langle \mathcal{E_{\alpha }}(s)|\frac{d}{ds}|%
\mathcal{D_{\alpha }}(s)\rangle \rangle 
\end{equation}%
and $V_{\beta \alpha }(s)$ by 
\begin{equation}
V_{\beta \alpha }(s)=p_{\beta }(s)\,\langle \langle \mathcal{E_{\alpha }}(s)|%
\frac{d\mathcal{L}(s)}{ds}|\mathcal{D_{\beta }}(s)\rangle \rangle .
\label{vba}
\end{equation}%
The integrand in the last line of Eq.~(\ref{scint}) can be rearranged in a
similar way to Eq.~(\ref{ricc}) for the closed case, yielding 
\begin{eqnarray}
&&\frac{V_{\beta \alpha }(s)}{\omega _{\beta \alpha }(s)}%
\,e^{T\,\Omega _{\beta \alpha }(s)}  \nonumber \\
&=&\frac{1}{T}\left[ \frac{d}{ds}\left( \frac{V_{\beta \alpha }}{\omega
_{\beta \alpha }^{2}}\,e^{T\,\Omega _{\beta \alpha }(s)}\right)
-e^{T\,\Omega _{\beta \alpha }(s)}\frac{d}{ds}\frac{V_{\beta \alpha }}{%
\omega _{\beta \alpha }^{2}}\right] .  \label{ninte}
\end{eqnarray}%
Therefore, from Eq.~(\ref{scint}) we have 
\begin{eqnarray}
&&p_{\alpha }(s)\,=\,p_{\alpha }(0)-\int_{0}^{s}ds^{\prime
}p_{\alpha }(s^{\prime })\,\Phi _{\alpha }(s^{\prime })  \nonumber \\
&&+\frac{1}{T}\sum_{\beta \neq \alpha }\left( \frac{V_{\beta \alpha }(0)}{%
\omega _{\beta \alpha }^{2}(0)}-\frac{V_{\beta \alpha }(s)}{\omega _{\beta
\alpha }^{2}(s)}\,e^{T\,\Omega _{\beta \alpha }(s)}\right.   \nonumber \\
&&\left. +\int_{0}^{s}ds^{\prime }\,e^{T\,\Omega _{\beta \alpha }(s^{\prime
})}\frac{d}{ds^{\prime }}\frac{V_{\beta \alpha }(s^{\prime })}{\omega
    _{\beta \alpha }^{2}(s^{\prime })}\right) .
  \label{afsc}
\end{eqnarray}
Thus a condition for adiabaticity in terms of the total time of evolution
can be given by comparing $T$ to the terms involving indices $\beta \neq
\alpha $. This can be formalized as follows.


\begin{proposition}
\label{t2} Consider an open quantum system whose Lindblad superoperator $\mathcal{L}(s)$ 
has the following properties: $(a)$ The Jordan decomposition
of $\mathcal{L}(s)$ is given by one-dimensional blocks. $(b)$ Each
eigenvalue of $\mathcal{L}(s)$ is associated to a unique Jordan block. Then 
the adiabatic dynamics in the interval $0\leq s\leq 1$ occurs if and only if the following 
time conditions, obtained for each Jordan block $\alpha$ of $\mathcal{L}(s)$, are satisfied:
\begin{eqnarray}
T &\gg& \max_{0\leq s\leq 1} \left\vert \,\sum_{\beta \neq \alpha }\left( 
\frac{V_{\beta \alpha }(0)}{\omega _{\beta \alpha }^{2}(0)}-\frac{V_{\beta
\alpha }(s)}{\omega _{\beta \alpha }^{2}(s)}\,e^{T\,\Omega _{\beta \alpha
}(s)}\right. \right.  \nonumber \\
&&\left. \left. +\int_{0}^{s}ds^{\prime }\,e^{T\,\Omega _{\beta \alpha
}(s^{\prime })}\frac{d}{ds^{\prime }}\frac{V_{\beta \alpha }(s^{\prime })}{%
\omega _{\beta \alpha }^{2}(s^{\prime })}\right) \right\vert,
\label{eq:Tad}
\end{eqnarray}
\end{proposition}

Equation (\ref{eq:Tad}) simplifies in a number of situations.

\begin{itemize}
\item Adiabaticity is guaranteed whenever $V_{\beta \alpha }$ vanishes for
all $\alpha \neq \beta $. An example of this case will be provided in Sec. %
\ref{example}.

\item Adiabaticity is similarly guaranteed whenever $V_{\beta \alpha }(s)$,
which can depend on $T$ through $p_{\beta }$, vanishes for all $\alpha
,\beta $ such that $\mathrm{Re}(\Omega _{\beta \alpha })>0$ and does not
grow faster, as a function of $T$, than $\exp (T|\,{\mathrm{Re}}\Omega
_{\beta \alpha }|)$ for all $\alpha ,\beta $ such that $\mathrm{Re}(\Omega
_{\beta \alpha })<0$.

\item When $\mathrm{Re}(\Omega _{\beta \alpha })=0$ and $\mathrm{Im}(\Omega
_{\beta \alpha })\neq 0$ the integral in inequality (\ref{eq:Tad}) vanishes
in the infinite time limit due to the Riemann-Lebesgue lemma~\cite%
{Churchill:book}, as in the closed case discussed before. In this case,
again, adiabaticity is guaranteed provided $p_{\beta }(s)$ [and hence $%
V_{\beta \alpha }(s)$] does not diverge as a function of $T$ in the limit $T
\rightarrow \infty$.

\item When $\mathrm{Re}(\Omega _{\beta \alpha })>0$, the adiabatic regime
can still be reached for large $T$ provided that $p_{\beta }(s)$ contains a
decaying exponential which compensates for the growing exponential due to $%
\mathrm{Re}(\Omega _{\beta \alpha })$.

\item Even if there is an overall growing exponential in inequality (\ref%
{eq:Tad}), adiabaticity could take place over a finite time interval $%
[0,T_{\ast }]$ and, afterwards, disappear. In this case, which would be an
exclusive feature of open systems, the crossover time $T_{\ast }$ would be
determined by an inequality of the type $T\gg a+b\exp (cT)$, with $c>0$. The
coefficients $a,b$ and $c$ are functions of the system-bath interaction.
Whether the latter inequality can be solved clearly depends on the values of 
$a,b,c$, so that a conclusion about adiabaticity in this case is model
dependent.
\end{itemize}


\subsubsection{General Jordan blocks}

We show now that the hypotheses $(a)$ and $(b)$ can be
relaxed, providing a generalization of Proposition~\ref{t2} for the case of
multidimensional Jordan blocks and Lindblad eigenvalues associated to more
than one independent eigenvector. Let us redefine our general coefficient $%
r_{\alpha }^{(i)}(t)$ as 
\begin{equation}
r_{\alpha }^{(i)}(t)=p_{\alpha }^{(i)}(t)\,e^{\int_{0}^{t}\lambda _{\alpha
}(t^{\prime })dt^{\prime }},
\end{equation}%
which, applied to Eq.~(\ref{rfinal}), yields 
\begin{eqnarray}
&&{\dot{p}}_{\alpha }^{(i)}\,=\,p_{\alpha }^{(i+1)}  \nonumber
\\
&&-\sum_{\beta \,|\,\lambda _{\beta }=\lambda _{\alpha
}}\sum_{j=0}^{n_{\beta }-1}p_{\beta }^{(j)}\langle \langle \mathcal{E}%
_{\alpha }^{(i)}|{\dot{\mathcal{D}}}_{\beta }^{(j)}\rangle \rangle
\,e^{\Omega _{\beta \alpha }}  \nonumber \\
&&-\sum_{\beta \,|\,\lambda _{\beta }\neq \lambda _{\alpha
}}\sum_{j=0}^{n_{\beta }-1}p_{\beta }^{(j)}\langle \langle \mathcal{E}%
_{\alpha }^{(i)}|{\dot{\mathcal{D}}}_{\beta }^{(j)}\rangle \rangle
\,e^{\Omega _{\beta \alpha }}.  \label{rfscg}
\end{eqnarray}%
The above equation can be rewritten in terms of the scaled time $s=t/T$. The
integration of the resulting expression then reads 
\begin{eqnarray}
&&p_{\alpha }^{(i)}(s)\,=\,p_{\alpha
}^{(i)}(0)+T\int_{0}^{s}ds^{\prime }p_{\alpha }^{(i+1)}(s^{\prime }) 
\nonumber \\
&&\hspace{-0.9cm}-\sum_{\beta \,|\,\lambda _{\beta }=\lambda _{\alpha
}}\sum_{j}\int_{0}^{s}ds^{\prime }p_{\beta }^{(j)}(s^{\prime })\,\Phi
_{\beta \alpha }^{(ij)}(s^{\prime })\,e^{T\,\Omega _{\beta \alpha
}(s^{\prime })}  \nonumber \\
&&\hspace{-0.9cm}-\sum_{\beta \,|\,\lambda _{\beta }\neq \lambda _{\alpha
}}\sum_{j,p}\int_{0}^{s}ds^{\prime }\frac{(-1)^{S_{p}}\,V_{\beta \alpha
}^{(ijp)}(s^{\prime })}{\omega _{\beta \alpha }^{p+S_{p}}(s^{\prime })}%
\,e^{T\,\Omega _{\beta \alpha }(s^{\prime })},  \label{scintg}
\end{eqnarray}%
where use has been made of Eq.~(\ref{rdd}), with the sum over $j$ and $p$ in
the last line denoting 
\begin{equation}
\sum_{j,p}\equiv \sum_{j=0}^{n_{\beta }-1}\sum_{p=1}^{(n_{\alpha }-i)}\left(
\prod_{q=1}^{p}\sum_{k_{q}=0}^{(j-S_{q-1})}\right) .
\end{equation}%
The function $\Phi _{\beta \alpha }^{(ij)}(s)$ is defined by 
\begin{equation}
\Phi _{\beta \alpha }^{(ij)}(s)=\langle \langle \mathcal{E}_{\alpha
}^{(i)}(s)|\frac{d}{ds}|{\mathcal{D}}_{\beta }^{(j)}(s)\rangle \rangle ,
\label{phabij}
\end{equation}%
and $V_{\beta \alpha }^{(ijp)}(s)$ by 
\begin{equation}
V_{\beta \alpha }^{(ijp)}(s)=p_{\beta }^{(j)}(s)\langle \langle \mathcal{E}%
_{\alpha }^{(i+p-1)}(s)|\frac{d\mathcal{L}(s)}{ds}|\mathcal{D}_{\beta
}^{(j-S_{p})}(s)\rangle \rangle .\,  \label{vbapj}
\end{equation}%
The term $T\int_{0}^{s}ds^{\prime }p_{\alpha }^{(i+1)}(s^{\prime })$ in the
first line of Eq.~(\ref{scintg}), which was absent in the case of
one-dimensional Jordan blocks analyzed above, has no effect on adiabaticity,
since it does not cause any mixing of Jordan blocks. Therefore, the analysis
can proceed very similarly to the case of one-dimensional blocks. Rewriting
the integral in the last line of Eq.~(\ref{scintg}), as we have done in
Eqs.~(\ref{akfinal}) and~(\ref{afsc}), and imposing the absence of mixing of
the eigenvalues $\lambda _{\beta }\neq \lambda _{\alpha }$, i.e., the
negligibility of the last line of Eq.~(\ref{scintg}), we find the following
general theorem ensuring the adiabatic behavior of an open system.


\begin{theorem}
\label{t3} Consider an open quantum system governed by a Lindblad
superoperator $\mathcal{L}(s)$. 
Then the adiabatic dynamics in the interval $0\leq s\leq 1$ occurs if 
and only if the following 
time conditions, obtained for each coefficient $p_\alpha^{(i)}(s)$, 
are satisfied:

\begin{eqnarray}
T &\gg& \max_{0\leq s\leq 1} \left\vert \,\sum_{\beta \,|\,\lambda _{\beta
}\neq \lambda _{\alpha }}\sum_{j,p}\,(-1)^{S_{p}}\right.  \nonumber \\
&&\times \left[ \frac{V_{\beta \alpha }^{(ijp)}(0)}{\omega _{\beta \alpha
}^{p+S_{p}+1}(0)}-\frac{V_{\beta \alpha }^{(ijp)}(s)\,e^{T\,\Omega _{\beta
\alpha }(s)}}{\omega _{\beta \alpha }^{p+S_{p}+1}(s)}\right.  \nonumber \\
&&\left. \left. +\int_{0}^{s}ds^{\prime }\,e^{T\,\Omega _{\beta \alpha
}(s^{\prime })}\frac{d}{ds^{\prime }}\frac{V_{\beta \alpha
}^{(ijp)}(s^{\prime })}{\omega _{\beta \alpha }^{p+S_{p}+1}(s^{\prime })}%
\right] \right\vert .  \label{eq:tadscg}
\end{eqnarray}
\end{theorem}

Theorem \ref{t3} provides a very general condition for adiabaticity in open
quantum systems. The comments made about simplifying circumstances, in the
case of one-dimensional blocks above, hold here as well. Moreover, a simpler
sufficient condition can be derived from Eq. (\ref{eq:tadscg}) by
considering the term with maximum absolute value in the sum. This procedure
leads to the following corollary:

\begin{corollary}
\label{ct3} A sufficient time condition for the adiabatic regime of an open
quantum system governed by a Lindblad superoperator $\mathcal{L}(t)$ is 
\begin{eqnarray}
T &\gg& \mathcal{M}_{ij}^{n_\alpha n_\beta} \, \max_{0\le s\le 1} \left\vert
\, \frac{V_{\beta \alpha }^{(ijp)}(0)}{\omega _{\beta \alpha}^{p+S_{p}+1}(0)}
-\frac{V_{\beta \alpha }^{(ijp)}(s)\,e^{T\,\Omega_{\beta \alpha }(s)}}{%
\omega _{\beta \alpha}^{p+S_{p}+1}(s)} \right.  \nonumber \\
&&\left.+\int_{0}^{s}ds^{\prime }\,e^{T\,\Omega _{\beta \alpha }(s^{\prime
})}\frac{d}{ds^{\prime }} \frac{V_{\beta \alpha }^{(ijp)}(s^{\prime })}{%
\omega _{\beta\alpha }^{p+S_{p}+1}(s^{\prime })} \right\vert,
\label{eq:tadcol}
\end{eqnarray}
where $\max $ is taken over all possible values of the indices $%
\lambda_\alpha \neq \lambda_\beta $, $i$, $j$, and $p$, with 
\begin{eqnarray}
&&\mathcal{M}_{ij}^{n_\alpha n_\beta} = \sum_{\beta \,|\,\lambda _{\beta
}\neq \lambda _{\alpha}} \sum_{j=0}^{(n_{\beta}-1)}\sum_{p=1}^{(n_{\alpha
}-i)}\left( \prod_{q=1}^{p}\sum_{k_{q}=0}^{(j-S_{q-1})}\right) 1  \nonumber
\\
&&= \Lambda_{\beta\alpha} \left[ \frac{(n_\alpha+n_\beta-i+1)!}{%
(n_\alpha-i+1)!n_\beta!}-n_\beta-1 \right],  \label{Nlt}
\end{eqnarray}
where $\Lambda_{\beta\alpha}$ denotes the number of Jordan blocks such that $%
\lambda_\alpha \neq \lambda_\beta$.
\end{corollary}




\subsection{Physical interpretation of the adiabaticity condition}


There are various equivalent ways in which to interpret the adiabatic
theorem for \emph{closed} quantum systems~\cite{Messiah:book}. A
particularly useful interpretation follows from Eq.~(\ref{timead2}): the
evolution time must be much longer than the ratio of the norm of the
time derivative of the Hamiltonian to the square of the spectral gap. In
other words, either the Hamiltonian changes slowly, or the spectral gap is
large, or both. It is tempting to interpret our results in a similar
fashion, which we now do.

The quantity $V_{\beta \alpha }^{(ijp)}$, by Eq.~(\ref{vbapj}), plays the
role of the time derivative of the Lindblad superoperator. However, the
appearance of $\exp [T\,\mathrm{Re}\,\Omega _{\beta \alpha }(s)]$ in Eq. (%
\ref{eq:tadscg}) has no analog in the closed-systems case, because the
eigenvalues of the Hamiltonian are real, while in the open-systems case the
eigenvalues of the Lindblad superoperator may have imaginary parts. This
implies that adiabaticity is a phenomenon which is not guaranteed to happen
in open systems even for very slowly varying interactions. Indeed, from
Theorems \ref{t2} and \ref{t3}, possible pictures of such system evolutions
include the decoupling of Jordan blocks only over a finite time interval
(disappearing afterwards), or even the case of complete absence of
decoupling for any time $T$, which implies no adiabatic evolution whatsoever.

The quantity $\omega _{\beta \alpha}$, by Eq.~(\ref{eq:omab}), clearly plays
the role of the spectral gap in the open-system case. There are two
noteworthy differences compared to the closed-system case. First, the $%
\omega _{\beta \alpha }$ can be complex. This implies that the differences
in decay rates, and not just in energies, play a role in determining the
relevant gap for open systems. Second, for multidimensional Jordan blocks, 
the terms $\omega_{\beta\alpha}$ depend on distinct powers for distinct pairs $\beta,\alpha$. 
Thus certain $\omega _{\beta \alpha }$ (those with the higher exponents) will
play a more dominant role than others. 

The conditions for adiabaticity are best illustrated further via examples,
one of which we provide next.



\section{Example: The adiabatic evolution of an open quantum two-level system%
}

\label{example}

\begin{figure}[h]
\centering
{\includegraphics[angle=0,scale=0.32]{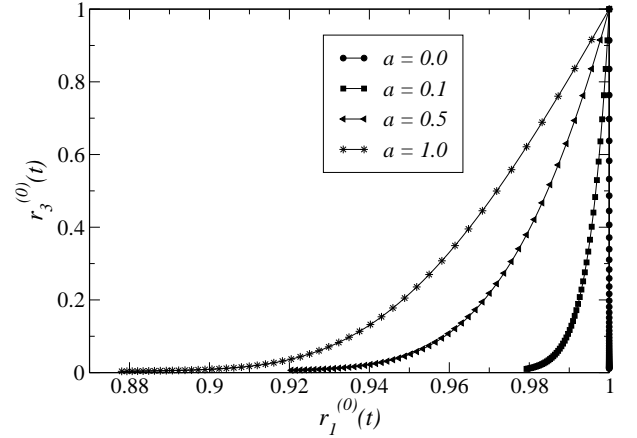}}
\caption{Parametric evolution of the coefficients $r_{3}^{(0)}(t)$ and 
$r_{1}^{(0)}(t)$ for $0\le t \le 1$. The initial conditions are 
$r_{1}^{(0)}(0)=r_{3}^{(0)}(0)=1.0$ and the decoherence parameters 
are taken as linear functions of time, i.e., 
$\epsilon (t)=\epsilon_{0}+at$ and $\gamma (t)=\gamma_{0}+at$, with 
$\epsilon_0=1.0$ and $\gamma_0=0.5$. The
master equation is solved numerically for $a>0$. 
In the adiabatic regime, corresponding to $a=0$, the evolution of  
$r_{1}^{(0)}(t)$ and $r_{3}^{(0)}(t)$ is decoupled, with 
$r_{1}^{(0)}(t) = 1$ independently of the value of $r_{3}^{(0)}(t)$.}
\label{f1}
\end{figure}

In order to illustrate the consequences of open quantum system adiabatic
dynamics, let us consider a concrete example that is analytically solvable.
Suppose a quantum two-level system, with internal Hamiltonian $H=(\omega
/2)\,\sigma _{z}$, and described by the master equation (\ref{eq:t-Lind}),
is subjected to two sources of decoherence: spontaneous emission $\Gamma
_{1}(t)=\epsilon (t)\,\sigma _{-}$ and bit flips $\Gamma _{2}(t)=\gamma
(t)\,\sigma _{x}$, where $\sigma _{-}=\sigma _{x}-i\sigma _{y}$ is the
lowering operator. Writing the density operator in the basis $\left\{
I_{2},\sigma _{x},\sigma _{y},\sigma _{z}\right\} $, i.e., as $\rho =(I_{2}+%
\overrightarrow{v}\cdot \overrightarrow{\sigma })/2$, Eq.~(\ref{le}) results
in 
\begin{equation}
|{\dot{\rho}}(t)\rangle \rangle =\frac{1}{2}\left( 
\begin{array}{c}
0 \\ 
-\omega v_{y}-2\epsilon ^{2}v_{x} \\ 
\omega v_{x}-2(\gamma ^{2}+\epsilon ^{2})v_{y} \\ 
-4\epsilon ^{2}-2(\gamma ^{2}+2\epsilon ^{2})v_{z} \\ 
\end{array}%
\right) =\frac{1}{2}\left( 
\begin{array}{c}
0 \\ 
{\dot{v}}_{x} \\ 
{\dot{v}}_{y} \\ 
{\dot{v}}_{z} \\ 
\end{array}%
\right) ,  \label{rhoex}
\end{equation}%
where $v_{x}(t)$, $v_{y}(t)$, and $v_{z}(t)$ are real functions providing
the coordinates of the quantum state $|\rho (t)\rangle \rangle $ on the
Bloch sphere. The Lindblad superoperator is then given by 
\begin{equation}
\mathcal{L}(t)=\left( 
\begin{array}{cccc}
0 & 0 & 0 & 0 \\ 
0 & -2\,\epsilon ^{2} & -\omega  & 0 \\ 
0 & \omega  & -2\epsilon ^{2}-2\gamma ^{2} & 0 \\ 
-4\,\epsilon ^{2} & 0 & 0 & -4\epsilon ^{2}-2\gamma ^{2}%
\end{array}%
\right) .  \label{lmat}
\end{equation}%
In order to exhibit an example that has a nontrivial Jordan block structure,
we now assume $\gamma ^{2}=\omega $ (which can in practice be obtained by
measuring the relaxation rate $\gamma $ and correspondingly adjusting the
system frequency $\omega $). We then have three different eigenvalues for $%
\mathcal{L}(t)$, 
\begin{eqnarray*}
\lambda _{1} &=&0, \\
\lambda _{2} &=&-2\epsilon ^{2}-\gamma ^{2}\text{ }\mathrm{(twofold\,\, degenerate)} \\
\lambda _{3} &=&-4\epsilon ^{2}-2\gamma ^{2},
\end{eqnarray*}%
which are associated with the following three independent (unnormalized)
right eigenvectors:
\begin{equation}
|\mathcal{D}_{1}^{(0)}\rangle \rangle =\left( 
\begin{array}{c}
f(\gamma ,\epsilon ) \\ 
0 \\ 
0 \\ 
1 \\ 
\end{array}%
\right) ,|\mathcal{D}_{2}^{(0)}\rangle \rangle =\left( 
\begin{array}{c}
0 \\ 
1 \\ 
1 \\ 
0 \\ 
\end{array}%
\right) ,|\mathcal{D}_{3}^{(0)}\rangle \rangle =\left( 
\begin{array}{c}
0 \\ 
0 \\ 
0 \\ 
1 \\ 
\end{array}%
\right) ,  \label{dex}
\end{equation}%
with $f(\gamma ,\epsilon )=-1-(\gamma ^{2}/2\epsilon ^{2})$. Similarly, for
the left eigenvectors, we find 
\begin{eqnarray}
\langle \langle \mathcal{E}_{1}^{(0)}| &=&\left( \frac{{}}{{}}1/f(\gamma
,\epsilon ),0,0,0\frac{{}}{{}}\right) ,  \nonumber \\
\vspace{0.1cm}\langle \langle \mathcal{E}_{2}^{(1)}| &=&\left( \frac{{}}{{}}%
0,\gamma ^{2},-\gamma ^{2},0\frac{{}}{{}}\right) ,  \nonumber \\
\vspace{0.1cm}\langle \langle \mathcal{E}_{3}^{(0)}| &=&\left( \frac{{}}{{}}%
-1/f(\gamma ,\epsilon ),0,0,1\frac{{}}{{}}\right) .  \label{rex}
\end{eqnarray}%
The Jordan form of $\mathcal{L}(t)$ can then be written as 
\begin{equation}
\mathcal{L}_{J}(t)=\left( 
\begin{array}{cccc}
0 & 0 & 0 & 0 \\ 
0 & -2\epsilon ^{2}-\gamma ^{2} & 1 & 0 \\ 
0 & 0 & -2\epsilon ^{2}-\gamma ^{2} & 0 \\ 
0 & 0 & 0 & -4\epsilon ^{2}-2\gamma ^{2}%
\end{array}%
\right) ,  \label{ljmat}
\end{equation}%
(observe the two-dimensional middle Jordan block), with the transformation
matrix leading to the Jordan form being 
\begin{equation}
S(t)=\left( 
\begin{array}{cccc}
f(\gamma ,\epsilon ) & 0 & 0 & 0 \\ 
0 & 1 & \gamma ^{-2} & 0 \\ 
0 & 1 & 0 & 0 \\ 
1 & 0 & 0 & 1%
\end{array}%
\right) .  \label{smat}
\end{equation}%
Note that, in our example, each eigenvalue of $\mathcal{L}(t)$ is associated
to a unique Jordan block, since we do not have more than one independent
eigenvector for each $\lambda _{\alpha }$. We then expect that the adiabatic
regime will be characterized by an evolution which can be decomposed by
single Jordan blocks. In order to show that this is indeed the case, let us
construct a right and left basis preserving the block structure. To this
end, we need to introduce a right and a left vector for the Jordan block
related to the eigenvalue $\lambda _{2}$. As in Eqs.~(\ref{dj}) and~(\ref{rj}%
), we define the additional states as 
\begin{equation}
|\mathcal{D}_{2}^{(1)}\rangle \rangle _{J}=\left( 
\begin{array}{c}
0 \\ 
0 \\ 
1 \\ 
0 \\ 
\end{array}%
\right) ,\,\,\,_{J}\langle \langle \mathcal{E}_{2}^{(0)}|=\left( \frac{{}}{{}%
}0,1,0,0\frac{{}}{{}}\right) .  \label{drjex}
\end{equation}%
We then obtain, after applying the transformations $|\mathcal{D}%
_{2}^{(1)}(t)\rangle \rangle =S(t)\,|\mathcal{D}_{2}^{(1)}\rangle \rangle
_{J}$ and $\langle \langle \mathcal{E}_{2}^{(0)}(t)|=\,_{J}\langle \langle 
\mathcal{E}_{2}^{(0)}|\,S^{-1}(t)$, the right and left vectors 
\begin{equation}
|\mathcal{D}_{2}^{(1)}\rangle \rangle =\left( 
\begin{array}{c}
0 \\ 
\gamma ^{-2} \\ 
0 \\ 
0 \\ 
\end{array}%
\right) ,\,\,\,\langle \langle \mathcal{E}_{2}^{(0)}|=\left( \frac{{}}{{}}%
0,0,1,0\frac{{}}{{}}\right) .  \label{drex}
\end{equation}%
Expanding the coherence vector in the basis $\left\{ |\mathcal{D}_{\alpha
}^{(j)}(t)\rangle \rangle \right\} $, as in Eq.~(\ref{rtime}), the master
equation~(\ref{le}) yields 
\begin{eqnarray}
&&\hspace{-0.3cm}f(\gamma ,\epsilon )\,{\dot{r}}_{1}^{(0)}+{\dot{f}}(\gamma
,\epsilon )\,r_{1}^{(0)}=0,  \nonumber \\
&&\hspace{-0.3cm}{\dot{r}}_{2}^{(0)}-2\frac{{\dot{\gamma}}}{\gamma ^{3}}%
r_{2}^{(1)}+\frac{{\dot{r}}_{2}^{(1)}}{\gamma ^{2}}=-\left( 2\epsilon
^{2}+\gamma ^{2}\right) r_{2}^{(0)}-2\frac{\epsilon ^{2}}{\gamma ^{2}}%
r_{2}^{(1)},  \nonumber \\
&&\hspace{-0.3cm}{\dot{r}}_{2}^{(0)}=r_{2}^{(1)}-\left( 2\epsilon
^{2}+\gamma ^{2}\right) r_{2}^{(0)},  \nonumber \\
&&\hspace{-0.3cm}{\dot{r}}_{1}^{(0)}+{\dot{r}}_{3}^{(0)}=\left( -4\epsilon
^{2}-2\gamma ^{2}\right) \,r_{3}^{(0)},  \label{glee}
\end{eqnarray}%
It is immediately apparent from Eq.~(\ref{glee}) that the block related to
the eigenvalue $\lambda _{2}$ is already decoupled from the rest. On the
other hand, by virtue of the last equation, the blocks associated to $%
\lambda _{1}$ and $\lambda _{3}$ are coupled, implying a mixing in the
evolution of the coefficients $r_{1}^{(0)}(t)$ and $r_{3}^{(0)}(t)$. The
role of the adiabaticity will then be the suppression of this coupling. We
note that in this simple example, the coupling between $r_{1}^{(0)}(t)$ and $%
r_{3}^{(0)}(t)$ would in fact also be eliminated by imposing the probability
conservation condition $\mathrm{Tr}\rho =1$. However, in order to discuss
the effects of the adiabatic regime, let us permit a general time evolution
of all coefficients (i.e., probability \textquotedblleft
leakage\textquotedblright ) and analyze the adiabatic constraints. The
validity condition for adiabatic dynamics, given by Eq.~(\ref{vch}), yields 
\begin{equation}
\left\vert \frac{\langle \langle \mathcal{E}_{3}^{(0)}|\,{\dot{\mathcal{L}}}%
\,|\mathcal{D}_{1}^{(0)}\rangle \rangle }{\lambda _{1}-\lambda _{3}}%
\right\vert =\left\vert \frac{2\gamma ^{2}{\dot{\epsilon}}/\epsilon -2\gamma 
  {\dot{\gamma}}}{\gamma ^{2}+2\epsilon ^{2}}\right\vert \ll 
\left\vert \lambda_{1}-\lambda_{3} \right\vert.
\label{exvc}
\end{equation}
We first note that we have here the possibility of an adiabatic evolution
even without ${\dot{\mathcal{L}}}(t)\approx 0$ in general (i.e., for all its
matrix elements). Indeed, solving $\gamma ^{2}{\dot{\epsilon}}/\epsilon
=\gamma {\dot{\gamma}}$, Eq.~(\ref{exvc}) implies that independent evolution
in Jordan blocks will occur for $\epsilon (t)\propto \gamma (t)$. \ Since $%
f(\gamma ,\epsilon )=-1-(\gamma ^{2}/2\epsilon ^{2})$ is then constant in
time, it follows, from Eq.~(\ref{glee}), that $r_{1}^{(0)}(t)$ is constant in
time, which in turn ensures the decoupling of $r_{1}^{(0)}(t)$ and $%
r_{3}^{(0)}(t)$. In this case, it is a \emph{dynamical symmetry} (constancy
of the ratio of magnitudes of the spontaneous emission and bit-flip
processes), rather than the general slowness of ${\dot{\mathcal{L}}}(t)$, that is
responsible for the adiabatic behavior. The same conclusion is also obtained from the 
adiabatic condition~(\ref{vc}). Of course, Eq.~(\ref{exvc}) is automatically satisfied if $\mathcal{L}(t)$
is slowly varying in time, which means ${\dot{\gamma}}(t)\approx 0$ and ${%
\dot{\epsilon}}(t)\approx 0$. Assuming this last case, the following
solution is found: 
\begin{eqnarray}
&&r_{1}^{(0)}(t)=r_{1}^{(0)}(0),  \nonumber \\
&&r_{2}^{(0)}(t)=\left[ r_{2}^{(1)}(0)\,t+r_{2}^{(0)}(0)\right] \,{e}%
^{(-2\epsilon ^{2}-\gamma ^{2})\,t},  \nonumber \\
&&r_{2}^{(1)}(t)=r_{2}^{(1)}(0)\,{e}^{(-2\epsilon ^{2}-\gamma ^{2})\,t}, 
\nonumber \\
&&r_{3}^{(0)}(t)=r_{3}^{(0)}(0)\,{e}^{(-4\epsilon ^{2}-2\gamma ^{2})\,t}.
\label{solex}
\end{eqnarray}%
It is clear that the evolution is independent in the three distinct Jordan
blocks, with functions $r_{\alpha }^{(i)}(t)$ belonging to different sectors
evolving separately. The only mixing is between $r_{2}^{(0)}(t)$ and $%
r_{2}^{(1)}(t)$, which are components of the the same block. The decoupling 
of the coefficients $r_{1}^{(0)}(t)$ and $r_{3}^{(0)}(t)$ in the adiabatic 
limit is exhibited in Fig.~\ref{f1}. Observe that the adiabatic behavior is
recovered as the dependence of $\epsilon (t)$ and $\gamma (t)$ on $t$
becomes negligible. 

The original coefficients $v_{x}$, $v_{y}$, and $v_{z}$ in the Bloch sphere
basis $\left\{ I_{2},\sigma _{x},\sigma _{y},\sigma _{z}\right\} $ can be
written as combinations of the functions $r_{\alpha }^{(i)}$. Equation~(\ref%
{solex}) yields 
\begin{eqnarray}
v_{x}(t) &=&\left( v_{x}(0)+(v_{x}(0)-v_{y}(0))\gamma ^{2}\,t\right)
\,e^{(-2\epsilon ^{2}-\gamma ^{2})\,t},  \nonumber \\
v_{y}(t) &=&\left( v_{y}(0)+(v_{x}(0)-v_{y}(0))\gamma ^{2}\,t\right)
\,e^{(-2\epsilon ^{2}-\gamma ^{2})\,t},  \nonumber \\
v_{z}(t) &=&\left( v_{z}(0)-\frac{1}{f(\gamma ,\epsilon )}\right)
e^{(-4\epsilon ^{2}-2\gamma ^{2})\,t}+\frac{1}{f(\gamma ,\epsilon )}
\label{evs}
\end{eqnarray}%
with the initial conditions 
\begin{eqnarray}
&&v_{x}(0)=r_{2}^{(0)}(0)+\gamma ^{-2}r_{2}^{(1)}(0),  \nonumber \\
&&v_{y}(0)=r_{2}^{(0)}(0),  \nonumber \\
&&v_{z}(0)=\frac{1}{f(\gamma ,\epsilon )}+r_{3}^{(0)}(0),
\end{eqnarray}%
where now $r_{1}^{(0)}(0)=1/f(\gamma ,\epsilon )$ has been imposed in order
to satisfy the $\mathrm{Tr}\rho =1$ normalization condition. The Bloch
sphere is then characterized by an asymptotic decay of the Bloch coordinates 
$v_{x}$ and $v_{y}$, with $v_{z}$ approaching the constant value $1/f(\gamma
,\epsilon )$.

Finally, let us comment on the analysis of adiabaticity in terms of the
conditions derived in Sec.~\ref{sec:tot-t} for the total time of
evolution. Looking at the matrix elements of ${\dot {\cal L}} (t)$, it can be shown that, 
for $\beta\ne \alpha$, the only term $V_{\beta \alpha }^{(ijp)}$  
defined by Eq.~(\ref{vbapj}) which can be \textit{a priori} nonvanishing is $V_{13}$. 
Therefore, we have to
consider the energy difference $\omega _{13}=4\epsilon ^{2}+2\gamma ^{2}$.
Assuming that the decoherence parameters $\epsilon $ and $\gamma $ are
nonvanishing, we have $\omega _{13}>0$ and hence $\Omega _{13}>0$. This
signals the breakdown of adiabaticity,
unless $V_{13}=0$. However, as we saw above, $V_{13}\propto \langle \langle 
\mathcal{E}_{3}^{(0)}|\,{\dot{\mathcal{L}}}\,|\mathcal{D}_{1}^{(0)}\rangle
\rangle =2\gamma ^{2}{\dot{\epsilon}}/\epsilon -2\gamma {\dot{\gamma}}$ and
thus $V_{13}=0$ indeed implies the adiabaticity condition $\epsilon
(t)\propto \gamma (t)$, in agreement with the results obtained from Theorem~%
\ref{t1}. In this (dynamical symmetry) case adiabaticity holds exactly,
while if $\epsilon (t)$ is \emph{not} proportional to $\gamma (t)$, then there can be no adiabatic evolution. 
Thus, the present example, despite nicely illustrating our concept of adiabaticity in 
open systems, does not present us with the
opportunity to derive a nontrivial condition on $T$; such more general examples will
be discussed in a future publication.



\section{Conclusions and outlook}

\label{conclusions}


The concept of adiabatic dynamics is one of the pillars of the theory of
closed quantum systems. Here we have introduced its generalization to open
quantum systems. We have shown that under appropriate slowness conditions
the time-dependent Lindblad superoperator decomposes into dynamically
decoupled Jordan blocks, which are preserved under the adiabatic dynamics.
Our key results are summarized in Theorems \ref{t1} and \ref{t3}, which
state sufficient (and necessary in the case of Theorem \ref{t3}) conditions
for adiabaticity in open quantum systems. In particular, Theorem \ref{t3} also 
provides the condition for breakdown of the adiabatic evolution. This
feature has no analog in the more restricted case of closed quantum systems.
It follows here from the fact that the Jordan eigenvalues of the dynamical
superoperator -- the generalization of the real eigenvalues of a Hamiltonian
-- can have an imaginary part, which can lead to unavoidable transitions
between Jordan blocks. It is worth mentioning that all of our results have been 
derived considering systems exhibiting gaps in the Lindblad eigenvalue spectrum. 
It would be interesting to understand the notion of adiabaticity when no gaps 
are available, as similarly done for the closed case in Refs. \cite{Avron:98,Avron:99}. 
Moreover, two particularly intriguing applications of the
theory presented here are to the study of geometric phases in open systems
and to quantum adiabatic algorithms, both of which have received
considerable recent attention~\cite{Farhi:00,Farhi:01,Thomaz:03,Carollo:04,Sanders:04}.
We leave these as open problems for future research.


\begin{acknowledgments}
M.S.S. gratefully acknowledges the Brazilian agency CNPq for financial
support. D.A.L. gratefully acknowledges financial support from NSERC and the
Sloan Foundation. This material is partially based on research sponsored by
the Defense Advanced Research Projects Agency under the QuIST program and
managed by the Air Force Research Laboratory (AFOSR), under agreement
F49620-01-1-0468 (to D.A.L.).
\end{acknowledgments}


\end{document}